\documentclass[conference, 9pt]{IEEEtran}
\IEEEoverridecommandlockouts
\usepackage{cite}
\usepackage{amsmath,amssymb,amsfonts}
\usepackage{graphicx}
\usepackage{textcomp}
\usepackage{xcolor}
\usepackage{ dsfont }
\usepackage{booktabs}
\usepackage{tablefootnote}
\usepackage{pbox}
\usepackage{enumitem}
\usepackage{pifont}
\usepackage{tikz}
\usepackage{inconsolata}
\usepackage{algorithm}
\usepackage{algorithmicx}
\usepackage{lipsum}
\usepackage[noend]{algpseudocode}

\algnewcommand\algorithmicforeach{\textbf{for each}}
\algdef{S}[FOR]{ForEach}[1]{\algorithmicforeach\ #1\ \algorithmicdo}

\algnewcommand{\IIf}[1]{\State\algorithmicif\ #1\  \algorithmicthen}
\algnewcommand{\EndIIf}{\unskip}
\usepackage{mathrsfs}

\usepackage{multirow}
\usepackage{longtable}

\usepackage{hyperref}

\def\BibTeX{{\rm B\kern-.05em{\sc i\kern-.025em b}\kern-.08em
    T\kern-.1667em\lower.7ex\hbox{E}\kern-.125emX}}

\renewcommand{\IEEEbibitemsep}{0pt plus 0.5pt}
\makeatletter
\IEEEtriggercmd{\reset@font\normalfont\fontsize{6.9pt}{6.9pt}\selectfont}
\makeatother
\IEEEtriggeratref{1}

\renewcommand{\baselinestretch}{0.95}

\newcommand\encircle[1]{%
  \tikz[baseline=(X.base)] 
    \node (X) [draw, shape=circle, inner sep=-1, fill=black, text=white] {\strut #1};%
}

\usepackage{xspace}
\newcommand{\Design}{\textit{$\mathsf{CyberHD}$\xspace}}    

\begin{document}

\title{Late Breaking Results: Scalable and Efficient Hyperdimensional Computing for Network Intrusion Detection\vspace{-5mm}}

\author{
    \IEEEauthorblockN{Junyao Wang$^\dag$, Hanning Chen$^\dag$, Mariam Issa$^\dag$, Sitao Huang$^\S$, Mohsen Imani$^\dag$}
    \IEEEauthorblockN{\textit{$^\dag$ Department of Computer Science, University of California, Irvine, CA, United States}\\
    $^\S$ \textit{Department of Electrical Engineering and Computer Science, University of California, Irvine, CA, United States}
    \\\textit{\{junyaow4, hanningc, mariamai, sitaoh, m.imani\}}@uci.edu\vspace{-7mm}}
}


\maketitle

\begin{abstract}
Cybersecurity has emerged as a critical challenge for the industry. With the large complexity of the security landscape, sophisticated and costly deep learning models often fail to provide timely detection of cyber threats on edge devices. Brain-inspired hyperdimensional computing (HDC) has been introduced as a promising solution to address this issue. However, existing HDC approaches use static encoders and require very high dimensionality and hundreds of training iterations to achieve reasonable accuracy. This results in a serious loss of learning efficiency and causes huge latency for detecting attacks. In this paper, we propose $\Design$, an innovative HDC learning framework that identifies and regenerates insignificant dimensions to capture complicated patterns of cyber threats with remarkably lower dimensionality. Additionally, the holographic distribution of patterns in high dimensional space provides $\Design$ with notably high robustness against hardware errors. 
\end{abstract}

\section{Introduction}
\begin{figure}[!ht]
\centering
\includegraphics[width=\linewidth]{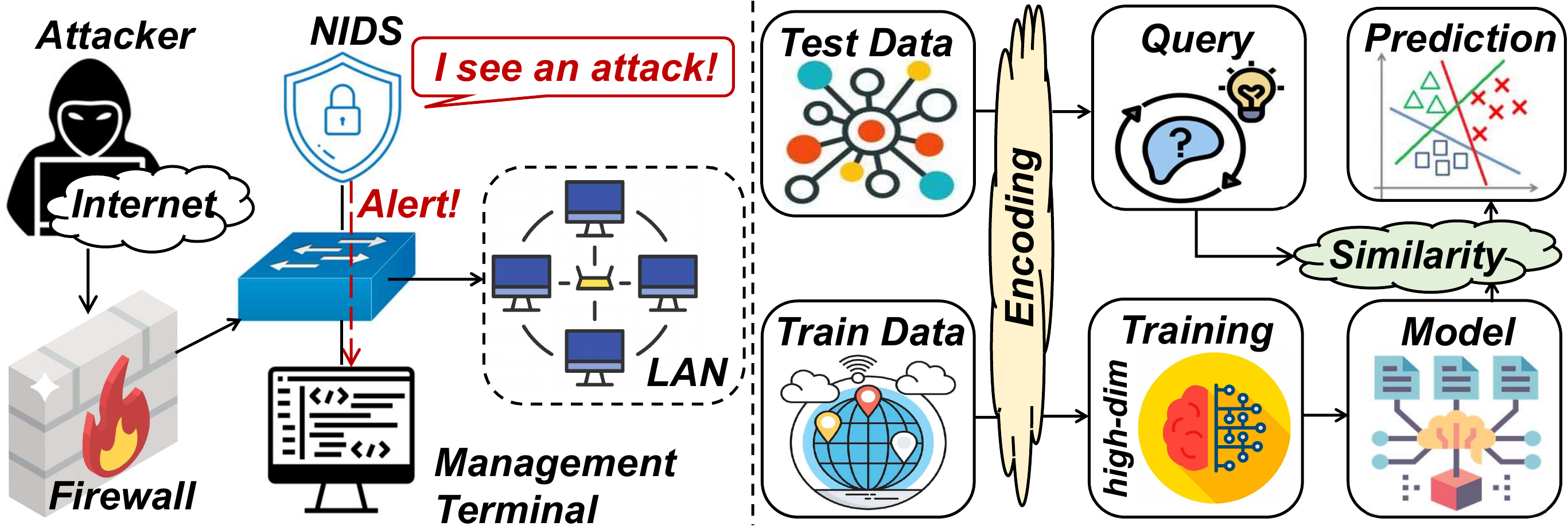}
\vspace{-6.5mm}
\caption{An Overview of NIDS and HDC Classification}
\vspace{-4mm}
\label{fig: background} 
\end{figure}

The rapid development of information technology has introduced increasingly sophisticated cyber threats. As one of the most widely deployed security devices, network intrusion detection systems (NIDS) is designed to monitor network traffic and identify suspicious activity. As shown in Fig. \ref{fig: background}(a), when traditional firewalls fail to intercept intruders, NIDS are aimed at providing timely detection and alerts to prevent the spread of infection through local area networks. With an increasingly gigantic amount of network traffic nowadays, along with the constant evolution of cyber attacks, a more intelligent, robust, and fully automated NIDS framework is of absolute necessity. 

Hyperdimensional Computing (HDC) is considered a promising solution for NIDS for its (\romannumeral 1) high computational efficiency ensuring real-time attack detection, (\romannumeral 2) holographical pattern distributions offering ultra-robustness against failures, and (\romannumeral 3) lightweight hardware implementations allowing efficient execution on edge devices~\cite{rahimi2016robust}. As shown in Fig. \ref{fig: background}(b), closely mimicking information representation and memorization functionalities of human brains, HDC encodes low-dimensional inputs to hypervectors with more than $10^4$ elements to perform various learning tasks. In this way, HDC conducts highly parallelizable operations and achieves high-quality results with significantly faster convergence and higher efficiency. 

Besides its lightweight and robust nature, utilizing encoded data points in hyperspace, HDC exhibits outstanding capabilities to distinguish various sophisticated attack patterns. Unfortunately, existing HDC algorithms use pre-generated encoders lacking the capability to distinguish the importance of each dimension, and hence require extremely high dimensionality to outperform DNNs. Consequently, the learning efficiency and scalability are compromised due to large numbers of unnecessary computations, especially for network intrusion detection tasks analyzing billions of network traffic instances. This work aims at addressing this issue by proposing $\Design$, a novel HDC learning framework utilizing a dynamic HDC encoding technique that identifies and regenerates insignificant dimensions. It ensures a highly effective dimensionality and achieves the desired accuracy with significantly higher efficiency.

\begin{figure}[!t]
\centering
\includegraphics[width=\linewidth]{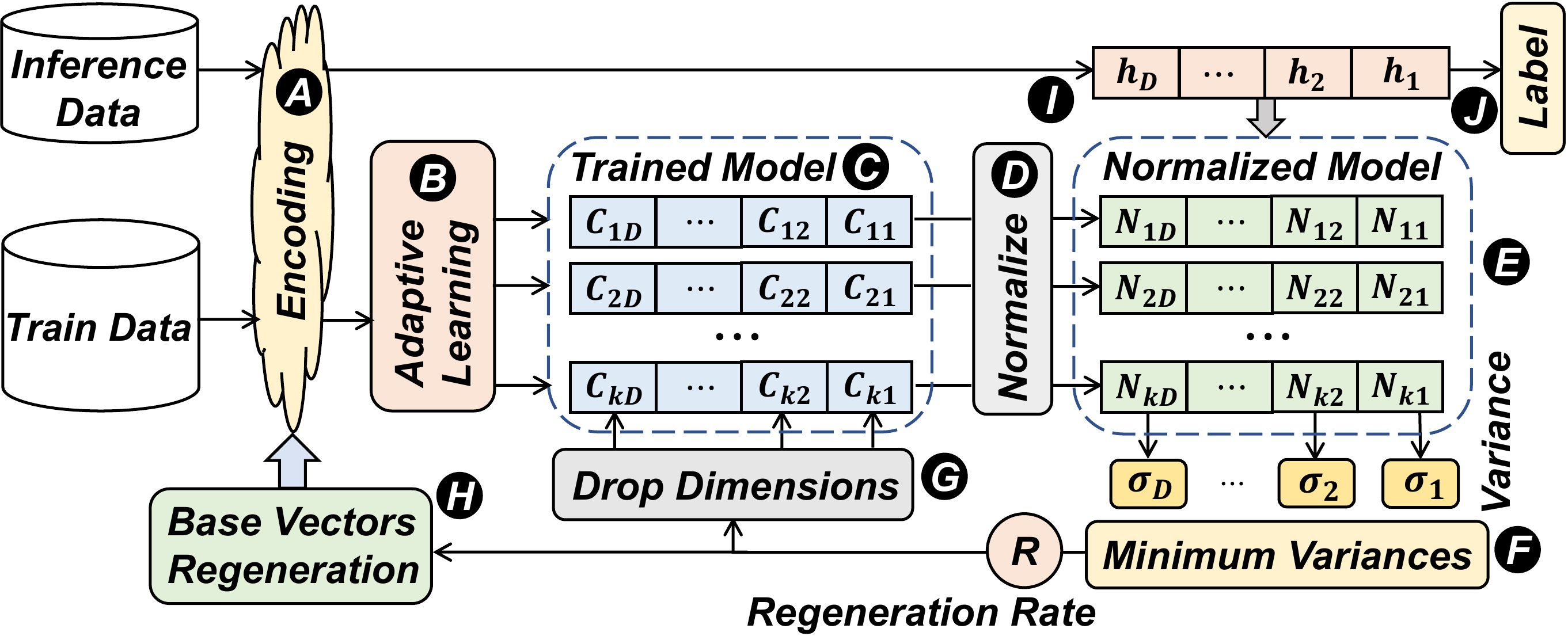}
\vspace{-6.5mm}
\caption{An Overview of $\Design$ Workflow}
\vspace{-3mm}
\label{fig: flow} 
\end{figure}

\begin{figure}[!ht]
\centering
\includegraphics[width=\linewidth]{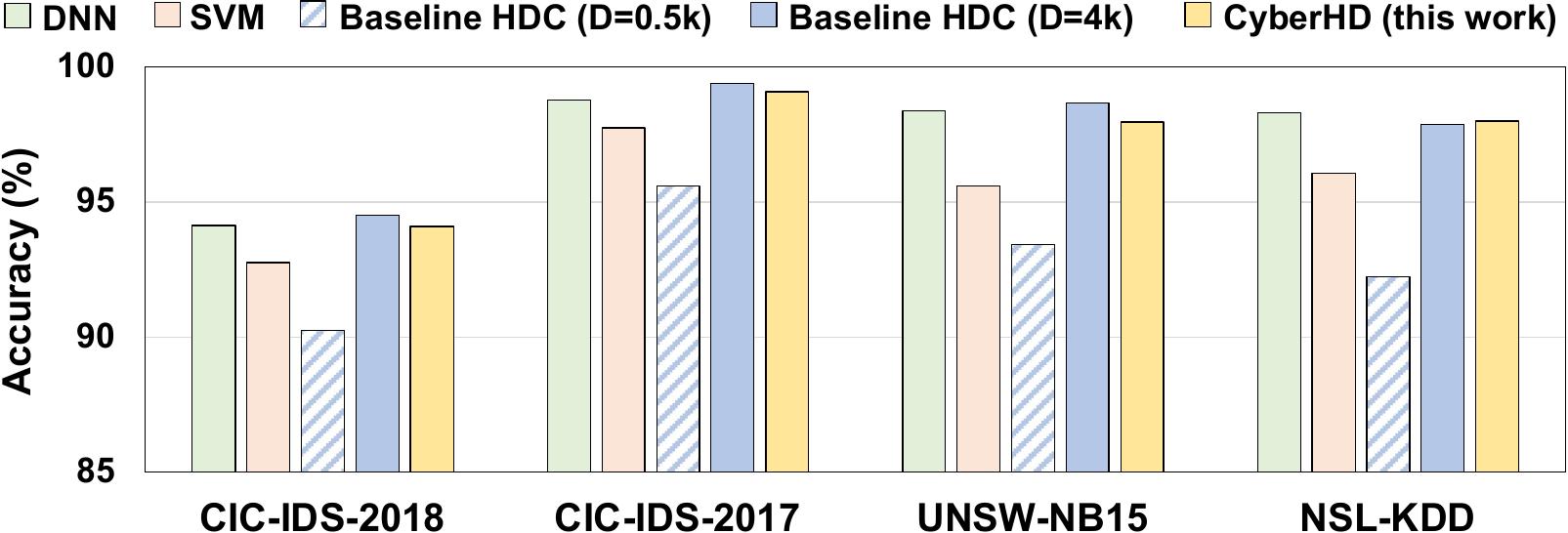}
\vspace{-6mm}
\caption{Comparing Accuracy of $\Design$ with State-of-the-art Algorithms}
\vspace{-6mm}
\label{fig: accuracy}  
\end{figure}

\begin{figure*}[!ht]
\vspace{-5mm}
  \centering
  {\includegraphics[width=\textwidth]{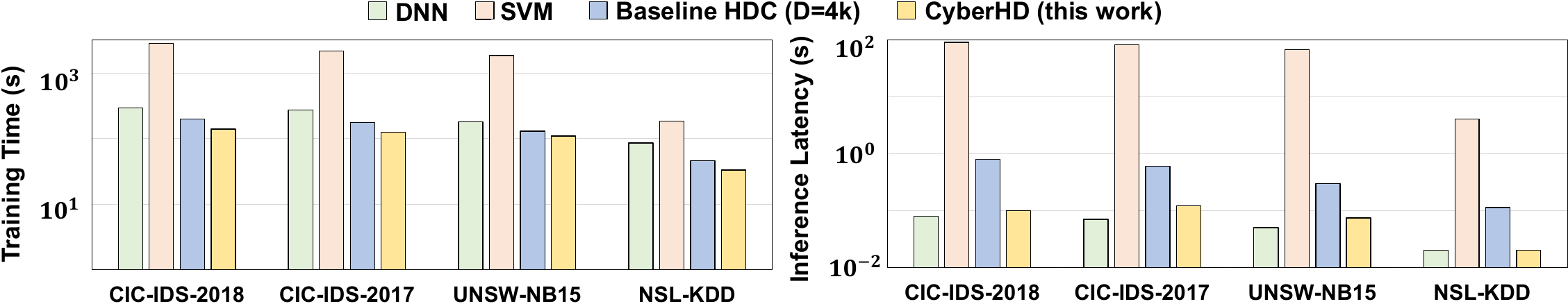}}
  \vspace{-6mm}
  \caption{\textbf{(log scale)} Comparing Training and Inference Efficiency of $\Design$ with SOTA Algorithms}
  \vspace{-6mm}
  \label{fig:time} 
\end{figure*}

\section{Related Work}
A recently proposed HDC framework, NeuralHD~\cite{zou2021scalable}, demonstrated a dynamic encoding approach by eliminating  dimensions with minor impacts on classifying patterns. 
Our work optimizes and parallelizes this dynamic HDC framework with matrix operations and applies it to the domain of cyber security for the first time, aiming to deliver real-time effective attack detection on resource-constrained devices. 

\section{Methodology} \label{sec:method}

As shown in Fig. \ref{fig: flow}, $\Design$ starts with encoding data points into high-dimensional space with existing encoding methods $(\encircle{A})$ depending on the data type. $\Design$ then leverage matrix operations to train the encoded data $(\encircle{B})$ in a highly-parallel way. We then normalize each class hypervector $(\encircle{D})$ and calculate the variance of each dimension over all classes  to identify dimensions with minimum impacts on the classification accuracy $(\encircle{F})$. $\Design$ then drops insignificant dimensions from our model depending on the regeneration rate $(\encircle{G})$. Finally, $\Design$ updates the base vectors on the selected dimensions and retrains the model$(\encircle{H})$.

\textbf{HDC Learning:} After encoding training data onto hyperspace $(\encircle{A})$, to reduce model saturation, we bundle encoded data by scaling a proper weight to each of them depending on how much new information they bring to class hypervectors. For instance, for a new encoded training sample $\mathcal H$, we update the model base on its cosine similarities with all class hypervectors, i.e. $\delta_l=\frac{ {\mathcal H} \cdot {\mathcal C_l}}{\| {\mathcal H}\|\cdot \| {\mathcal C_l}\|}$, where $ {\mathcal H} \cdot {\mathcal C_l}$ is the dot product between ${\mathcal H}$ and a class hypervector ${\mathcal C_l}$. 
If $\overrightarrow{\mathcal H}$ has the highest cosine similarity with class $l'$ while it actually has label $l$, the model updates as $\overrightarrow {\mathcal C_l} \leftarrow \overrightarrow {\mathcal C_l} + \eta(1-\delta_l)\times\overrightarrow{\mathcal H}$ and $\overrightarrow {\mathcal C_{l'}} \leftarrow \overrightarrow {\mathcal C_l} - \eta(1-\delta_{l'})\times\overrightarrow{\mathcal H}$, where $\eta$ is a learning rate. A large $\delta_l$ indicates the input data point is common or already exists in the model and updates the model by adding a very small portion of the encoded query ($1-\delta_l \approx 0$). 
In contrast, a small $\delta_l$ indicates a noticeably new pattern that is uncommon or does not already exist in the model so we update the model with a large factor ($1-\delta_l \approx 1$).

\textbf{Insignificant dimensions:} HDC represents each class with a class hypervector encoding patterns of that class$(\encircle{C})$. In inference, a query vector$(\encircle{I})$ is assigned to the class where it has the highest cosine similarity$(\encircle{J})$. An effective classifier achieves the desired accuracy with a strong capability to distinguish patterns so that query vectors have very differentiated cosine similarities to each class. In contrast, dimensions with similar values over all classes store common information and play minimal roles in differentiating patterns. $\Design$ identifies and drops such dimensions by calculating variances over all classes. After computing variances$(\encircle{F})$ over the normalized model $(\encircle{E})$, $\Design$ selects $\mathcal R\%$ of dimensions with the lowest variance to drop $(\encircle{G})$ depending on a regeneration rate $\mathcal R$. 

\textbf{Dimension Regeneration:}
To further improve classification accuracy, $\Design$ regenerates the dimensions selected to drop $(\encircle{H})$, so that the new dimensions can potentially better differentiate patterns. For cybersecurity datasets, considering the non-linear relationship between features, we utilize an encoder 
inspired by the Radial Basis Function (RBF)\cite{rahimi2007random}. During regeneration, $\Design$ replaces the base vector of the selected dimension in the encoding module with another randomly generated vector from Gaussian Distribution.


\section{Experimental Result} \label{sec:eval}
We evaluate $\Design$ with CPU (Intel Core i9-12900) and FPGA (Xilinx Alveo U50) on popular cyber-security datasets: NSL-KDD~\cite{tavallaee2009detailed}, UNSW-15~\cite{moustafa2015unsw}, CIC-IDS-2017~\cite{sharafaldin2018toward}, and CIC-IDS-2018~\cite{leevy2020survey}. We compare $\Design$ with state-of-the-art (SOTA) DNNs~\cite{taud2018multilayer} and SVMs~\cite{hearst1998support}. We also compare $\Design$ with SOTA HDCs~\cite{rahimi2016robust} without the capability to regenerate dimensions (baselineHD), and report results in two dimensionality: physical dimensionality $(\mathcal D=0.5\textrm{k})$ of $\Design$, and effective dimensionality $(\mathcal D^* = 4\textrm{k})$ as $\Design$. The effective dimensionality is defined as the sum of the physical dimensionality with regenerated dimensions throughout the retraining. 

\textbf{Accuracy:} As shown in Fig. \ref{fig: accuracy}, $\Design$ provides comparable accuracy to SOTA DNNs, while on average a $1.63\%$ higher accuracy than SVMs. $\Design$ also shows on average $4.28\%$ higher accuracy than baselineHD $(\mathcal D=0.5\textrm{k})$. Additionally, $\Design$ delivers comparable accuracy to baselineHD using the same dimensionality as the effective dimension$(\mathcal D^*=4\textrm{k})$, indicating that $\Design$ is capable of providing comparably high classification accuracy as the SOTA HDC while using $8.0\times$ lower physical dimensionality.

\textbf{Efficiency:}  For fairness, we compare the training and inference efficiency of the SOTA DNN, SVM, baselineHD ($\mathcal D^*=4$k), and $\Design$ ($\mathcal D=0.5$k) as they achieve comparable accuracy as shown in Fig. \ref{fig: accuracy}. Since cybersecurity datasets generally include millions of samples, SVM algorithms take an extraordinarily long time for both training and inference. Additionally, as shown in Fig. \ref{fig:time}, $\Design$ delivers on average $2.47\times$ faster training than SOTA DNN. $\Design$ also provides on average $1.85\times$ faster training and $15.29\times$ faster inference compared to the baselineHD ($\mathcal D^*=4$k).

\begin{table}[!t]
\scriptsize
\caption{Impact of Bitwidth on CPU's and FPGA's Energy Efficiency}\vspace{-4mm}
\begin{center}
\begin{tabular}{l|cccccc} \toprule
\textbf{} & \textbf{32 bits}  & \textbf{16 bits}  & \textbf{8 bits}  & \textbf{4 bits}  & \textbf{2 bits} & \textbf{1 bit}\\
\midrule
\textbf{Effective \textit{D}} & 1.2k & 2.1k & 3.6k & 5.6k & 7.5k & 8.8k \\
\midrule
\textbf{CPU} & $6.6\times$ & $4.0\times$ & $2.4\times$ & $1.5\times$ & $1.2\times$ & \textbf{1.0}$\times$ \\
\textbf{FPGA}& $16\times$ & $24\times$ & $34\times$ & $31\times$ & $28\times$ & $26\times$  \\
\bottomrule
\end{tabular}
\end{center}
\label{tb:hardware_efficiency} \vspace{-8mm}
\end{table}

\textbf{Cross Platform and Quantization Evaluation:} TABLE~\ref{tb:hardware_efficiency} shows the impact of the hypervectors' dimensions and bitwidths on the training efficiency, normalized to the efficiency of 1-bit CPU implementation. HDC models on CPUs achieve higher efficiency with low dimensionality and high element bitwidth due to the limited parallelism on CPU. CPUs demonstrate more strength for high bitwidth data due to their high frequency and powerful arithmetic logic unit (ALU). We also implement $\Design$ on FPGA with different hypervector dimensions and element bitwidths for comparison. FPGA shows excellent energy efficiency improvement compared to CPU due to its high parallelism and low power consumption. On the Xilinx Alveo U50 FPGA board, the power consumption of the $\Design$ accelerator is less than 20 W under 200 MHz frequency.

\begin{figure}[!ht]\vspace{-3mm}
\centering
\includegraphics[width=\linewidth]{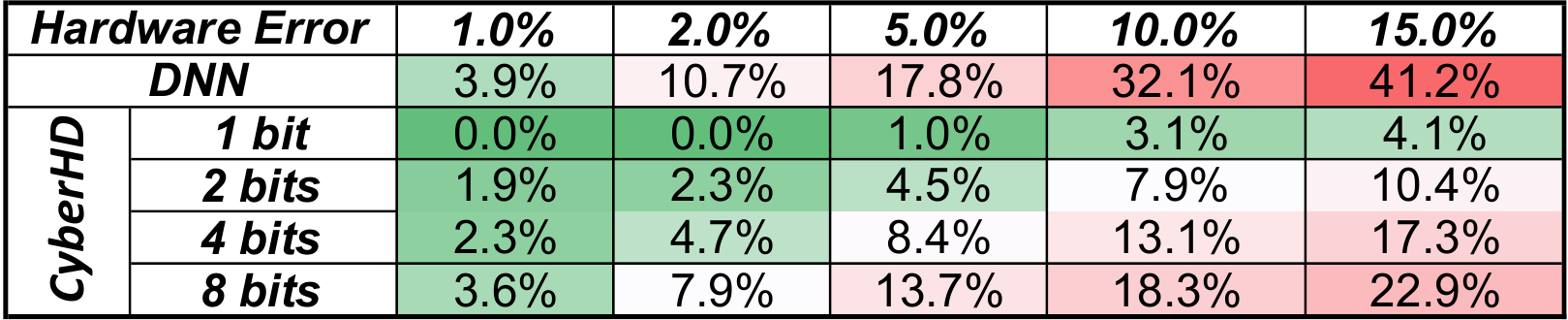}
\vspace{-6mm}
\caption{Comparing Robustness of $\Design$ with DNNs}
\vspace{-5mm}
\label{fig: robstness}  
\end{figure}

\textbf{Robustness Against Hardware Failures:} As shown in Fig. \ref{fig: robstness}, in DNNs, random bit flip results in significant quality loss as corruptions on most significant bits can cause major weight changes. In contrast, $\Design$ provides notably higher robustness against noise due to its redundant and holographic distribution. Additionally, all dimensions equally contribute to storing information so that failure on partial data will not result in the loss of entire information. $\Design$ delivers the maximum robustness using hypervectors in 1-bit precision, which is on average $12.90\times$ higher than the robustness of the DNN. An increase in precision lowers the robustness of $\Design$ since random flips on more significant bits will cause more accuracy loss. 

\section{Acknowledgements} \label{sec:conclusion}

This work was supported in part by National Science Foundation \#2127780, Semiconductor Research Corporation (SRC), Office of Naval Research, grants \#N00014-21-1-2225 and \#N00014-22-1-2067, the Air Force Office of Scientific Research under award \#FA9550-22-1-0253, and generous gifts from Xilinx and Cisco.

\bibliographystyle{unsrt}
\bibliography{reference}

\begin{thebibliography}{1}

\bibitem{rahimi2016robust}
Abbas Rahimi et~al.
\newblock A robust and energy-efficient classifier using brain-inspired
  hyperdimensional computing.
\newblock In {\em ISLPED}, 2016.

\bibitem{zou2021scalable}
Zhuowen Zou et~al.
\newblock Scalable edge-based hyperdimensional learning system with brain-like
  neural adaptation.
\newblock In {\em SC}, 2021.

\bibitem{rahimi2007random}
Ali Rahimi and Benjamin Recht.
\newblock Random features for large-scale kernel machines.
\newblock {\em Advances in neural information processing systems}, 2007.

\bibitem{tavallaee2009detailed}
Mahbod Tavallaee et~al.
\newblock {A detailed analysis of the KDD CUP 99 data set}.
\newblock In {\em CISDA}. IEEE, 2009.

\bibitem{moustafa2015unsw}
Nour Moustafa and Jill Slay.
\newblock {UNSW-NB15}: a comprehensive data set for network intrusion detection
  systems.
\newblock In {\em MilCIS}. IEEE, 2015.

\bibitem{sharafaldin2018toward}
Iman Sharafaldin et~al.
\newblock Toward generating a new intrusion detection dataset and intrusion
  traffic characterization.
\newblock {\em ICISSP}, 2018.

\bibitem{leevy2020survey}
Joffrey Leevy et~al.
\newblock {A survey and analysis of intrusion detection models based on
  CSE-CIC-IDS2018 big data}.
\newblock {\em Journal of Big Data}, 2020.

\bibitem{taud2018multilayer}
Hind Taud et~al.
\newblock Multilayer perceptron (mlp).
\newblock In {\em Geomatic approaches for modeling land change scenarios}.
  Springer, 2018.

\bibitem{hearst1998support}
Marti~A. Hearst et~al.
\newblock Support vector machines.
\newblock {\em Intelligent Systems and their applications}, 1998.

\end{thebibliography}
\end{document}